\documentclass[a4paper, 10pt, twocolumn]{article} 
\usepackage[british]{babel}
\usepackage{microtype}
\usepackage{amsmath,amsfonts,amsthm}
\usepackage[svgnames,table]{xcolor}
\definecolor{GammaDarkGreen}{rgb}{0.113,0.278,0.2745}
\definecolor{GammaCharcoal}{rgb}{0.0706, 0.0706, 0.0706}
\definecolor{GammaRed}{HTML}{B71325}
\definecolor{GammaGreen}{HTML}{004846}
\usepackage{booktabs}
\usepackage{lastpage}
\usepackage{graphicx}
\usepackage{enumitem}
\setlist{leftmargin=*,itemsep=0.4ex,parsep=0ex,topsep=0.8ex,partopsep=0ex}
\setlist[itemize]{label=\color{GammaRed}$\triangleright$}
\usepackage{sectsty}
\chapterfont{\normalfont\Large\color{GammaRed}}
\sectionfont{\normalfont\Large\color{GammaRed}}
\subsectionfont{\normalfont\large\color{GammaCharcoal}}
\subsubsectionfont{\normalfont\normalsize\color{GammaGreen}}

\usepackage{subfig}

\usepackage{caption}
\DeclareCaptionLabelFormat{abbrevfig}{\color{GammaRed}Fig.\ #2}
\DeclareCaptionLabelFormat{abbrevtab}{\color{GammaRed}Tab.\ #2}
\usepackage[pdftex]{hyperref}
\pdftrailerid{}
\hypersetup{colorlinks=true,linkcolor=GammaDarkGreen,citecolor=GammaDarkGreen,filecolor=GammaDarkGreen,urlcolor=GammaDarkGreen}

\addto\extrasenglish{%
}

\usepackage{geometry}
\usepackage[utf8]{inputenc}
\usepackage[T1]{fontenc}
\IfFileExists{nimbusmononarrow.sty}{\usepackage{nimbusmononarrow}}{}
\usepackage{tgheros}
\usepackage{sansmath}
\usepackage{fancyhdr}
\usepackage{titling}

\newcommand{\authorstyle}[1]{{\large\color{GammaCharcoal}#1}}
\newcommand{\institution}[1]{{\footnotesize\itshape\color{GammaDarkGreen}#1\/}}

\pretitle{%
    \noindent
    \begin{minipage}[t]{0.63\textwidth}\vspace{20pt}
        \LARGE\color{GammaRed}
}

\posttitle{%
    \end{minipage}%
    \hfill
    \begin{minipage}[t]{0.33\textwidth}\vspace{-50pt}
        \raggedleft
        \includegraphics[width=\linewidth]{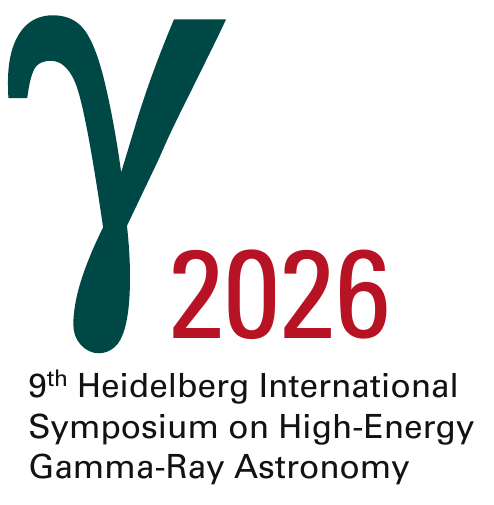}
    \end{minipage}%
}

\preauthor{\color{GammaDarkGreen}}
\postauthor{}
\predate{\begin{flushleft}\large\color{GammaRed}}
\postdate{\par\end{flushleft}}

\usepackage{lettrine}
\usepackage{xstring}
\newcommand{\initial}[1]{\lettrine[lines=2,findent=4pt,nindent=0pt]{\color{GammaRed}\emph{#1}}{}}
\newcommand{\lettrineabstract}[1]{\StrLeft{#1}{1}[\firstletter]\initial{\firstletter}\emph{{\StrGobbleLeft{#1}{1}}}}

\usepackage[backend=bibtex,style=numeric-comp,sorting=none,giveninits=true,citestyle=numeric]{biblatex}
\usepackage[autostyle=true]{csquotes}

\usepackage{aas_macros}

\newcommand{\pksf}{PKS~1510-089}
\newcommand{\pkst}{PKS~2155-304}
\newcommand{\g}{\ensuremath{\gamma}}

\title{
    A blazar is a blazar is a blazar \ldots \\ Comparing long-term varaibility properties of PKS~1510-089 and PKS~2155-304
}

\author{
	\authorstyle{M.~Zacharias\textcolor{GammaDarkGreen}{\textsuperscript{1}}, A.~Wierzcholska\textcolor{GammaDarkGreen}{\textsuperscript{2}}}
	\newline\newline
    {\footnotesize
        \textsuperscript{\color{GammaDarkGreen}1}\institution{Landessternwarte, Universit\"at Heidelberg, K\"onigstuhl 12, D-69117 Heidelberg, Germany}\\
        \textsuperscript{\color{GammaDarkGreen}2}\institution{Institute of Nuclear Physics, Polish Academy of Sciences, Radzikowskiego 152, 31-342 Krak\'{o}w, Poland}
    }\\
}

\date{\today}
\newcommand{\website}{\href{https://plan.events.mpg.de/event/543/}{Proceedings of the 9th Heidelberg Gamma-Ray Symposium}} 

\begin{document}
\sansmath\maketitle
\thispagestyle{fancy}

\lettrineabstract{Blazars are typically divided by spectral appearance; may that be strength of optical emission lines or the frequency of the synchrotron peak. However, this may say nothing about the variability properties of the various types. Here, we study the long-term variability properties with nearly 20 years of multiwavelength data of one flat-spectrum radio quasar, \pksf, and one high-frequency-peaked BL Lac object, \pkst. Both exhibit strong variability with an energy-dependent variability amplitude. Using the total data sets, we do not find any correlation between the various bands. While the flux distributions of \pkst\ are compatible with log-normal functions, this is not the case for \pksf. We conclude that the one-zone model is inadequate to describe the long-term evolution of blazars, requiring multi-zone or stratified models. Nevertheless, the underlying jet physics appear similar between the two types and in both sources the variability is, in fact, variable.
}

\section{Introduction}

Here, we summarize the results presented in detail in \cite{WZ26,ZW26}, which analysed the multiwavelength (MWL) data of \pksf\ and \pkst. Data were taken from \textit{Swift}-XRT and UVOT, as well as from \textit{Fermi}-LAT, spanning roughly 20 years from 2005 to 2024. The intention of analysing these vast data sets was to study the long-term variability properties of both sources, which are representatives of two different blazar subclasses -- of flat-spectrum radio quasars (FSRQs) and high-frequency-peaked BL Lac objects (HBLs).

Blazars are a subclass of active galaxies, where the relativistic jets points towards Earth \cite{blandfordrees74}. Blazars are categorized based on their spectral appearance. If broad optical emission lines are observed, the object is referred to as an FSRQ, while featureless optical spectra lead to the classification as a BL Lac object. The peak frequency of the low-energy hump in the spectral energy distribution (SED) also leads to subclassifications. Notably, if the synchrotron peak is located at $\nu>10^{15}\,$Hz, the objects are referred to as HBLs. The SED of blazars is dominated by two broad spectral components, where the low-energy one is synchrotron emission, while the high-energy one is likely inverse-Compton emission \cite{cerruti20} using either external photon fields (external Compton, EC) or synchrotron photons (synchrotron-self Compton, SSC). In some blazars, especially FSRQs, the accretion disk can also be prominent.

\begin{figure}
    \centering
    \includegraphics[width=0.48\textwidth]{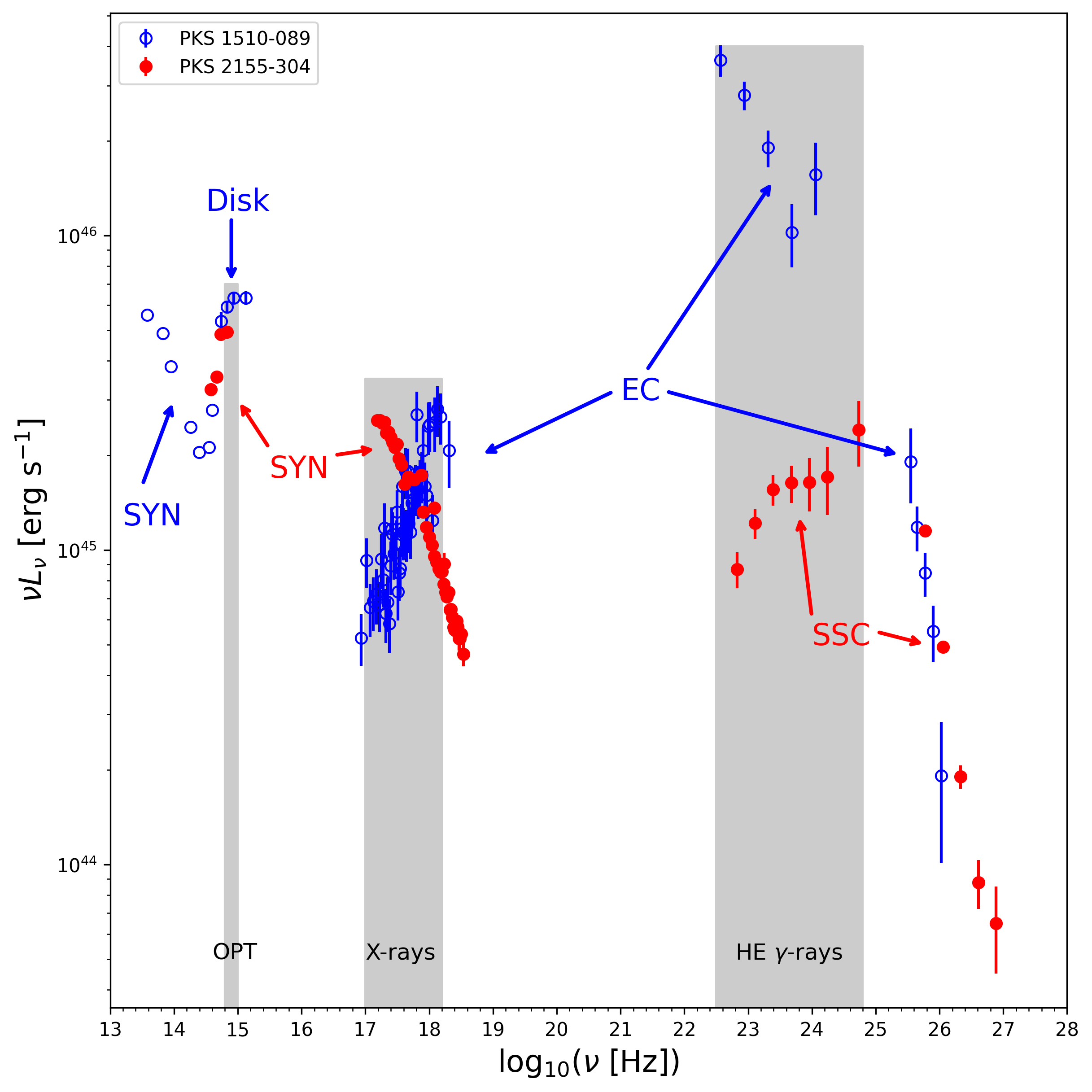}
    \caption{Average SEDs of \pksf\ \cite{hess23} and \pkst \cite{HESS_ATOM}. The gray shaded areas mark the considered energy bands, while the coloured text pieces depict the most likely radiative contributions.}
    \label{fig:SEDs}
\end{figure}

The aforedescribed SED features can be seen in Fig.~\ref{fig:SEDs}, where we show typical SEDs of the two sources we study here: \pksf, an FSRQ, and \pkst, an HBL. Both are quite famous sources for a variety of reasons including their persistency and variability at very-high-energy (VHE, $E>100\,$GeV) \g\ rays. \pksf\ is located at a redshift of $0.361$ and is the only FSRQ with persistent VHE flux \cite{magic18} potentially coming from significant distances from the black hole \cite{hess23}. \pkst\ is located at a redshift of $0.117$, and has seen massive X-ray and VHE \g-ray outbursts \cite{hess07pks,Aharonian2009_flare2}.

As stated before, we use 20 years of data from \textit{Swift} and \textit{Fermi} for both sources. The energy bands are marked in Fig.~\ref{fig:SEDs} as gray shaded regions. However, the spectral components of the blazars covered by the observed bands differ. The optical bands conver the high-energy part of the synchrotron component and a strong accretion disk component in \pksf, while they cover the low-energy synchrotron part in \pkst. The X-rays trace the low-energy part of the Compton component in \pksf, while it is the high energy part of the synchrotron component for \pkst. Lastly, the high-energy (HE, $E>100\,$MeV) \g\ rays observe the high-energy part of the Compton component in \pksf, while it is the low-energy part of the Compton component in \pkst. It is important, though, that the optical and HE \g\ rays should be produced by roughly the same electrons considering the usual one-zone model assumptions. We write these points down clearly, as it can become confusing in the discussion below. In the following, we show two examples of our studies -- namely, cross correlations and flux distributions -- while all details, including the full light curves, can be found in the dedicated papers \cite{WZ26,ZW26}.

\section{Cross Correlations}

\begin{figure*}
    \centering
    \includegraphics[width=0.48\textwidth]{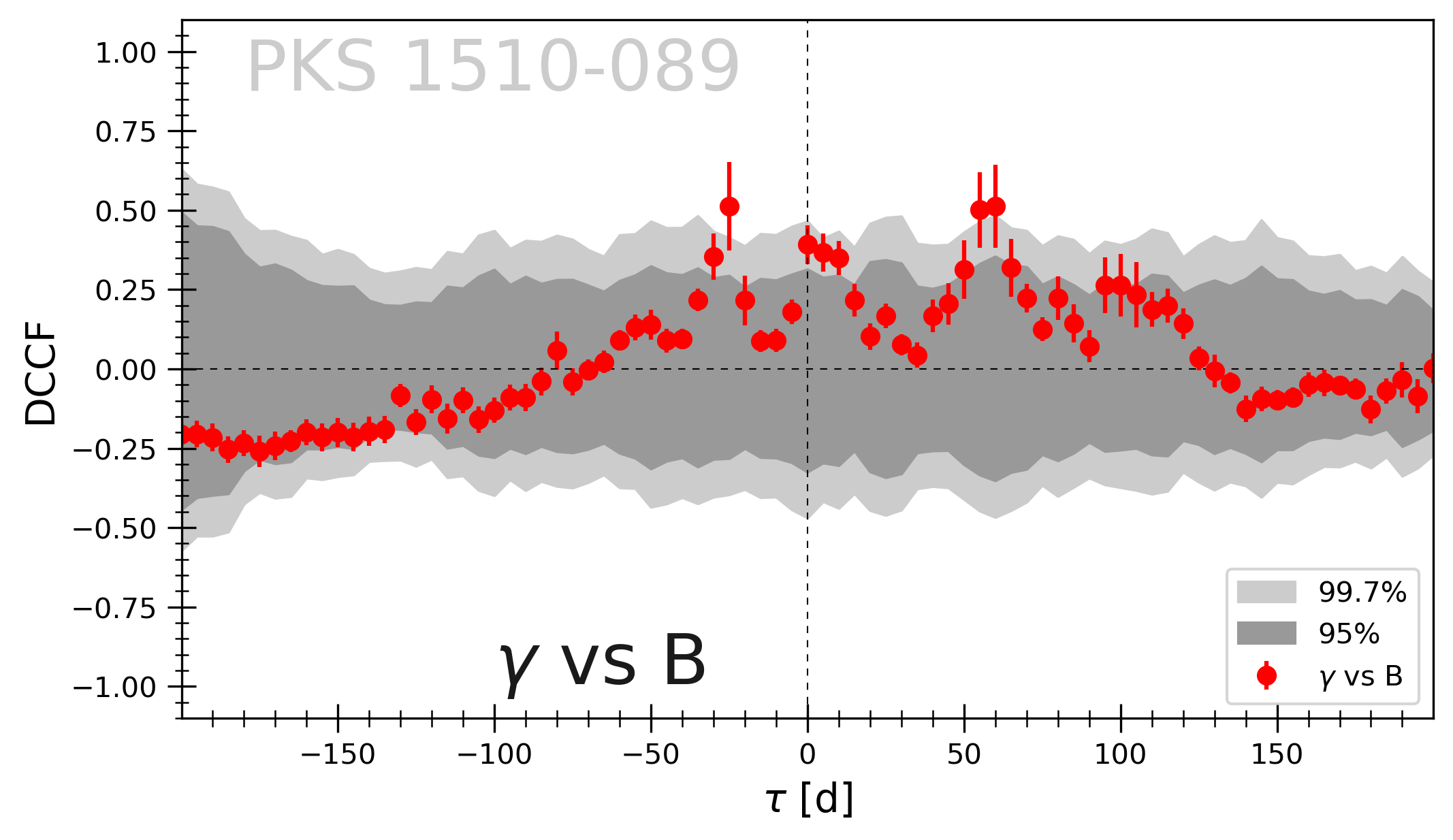}~
    \includegraphics[width=0.48\textwidth]{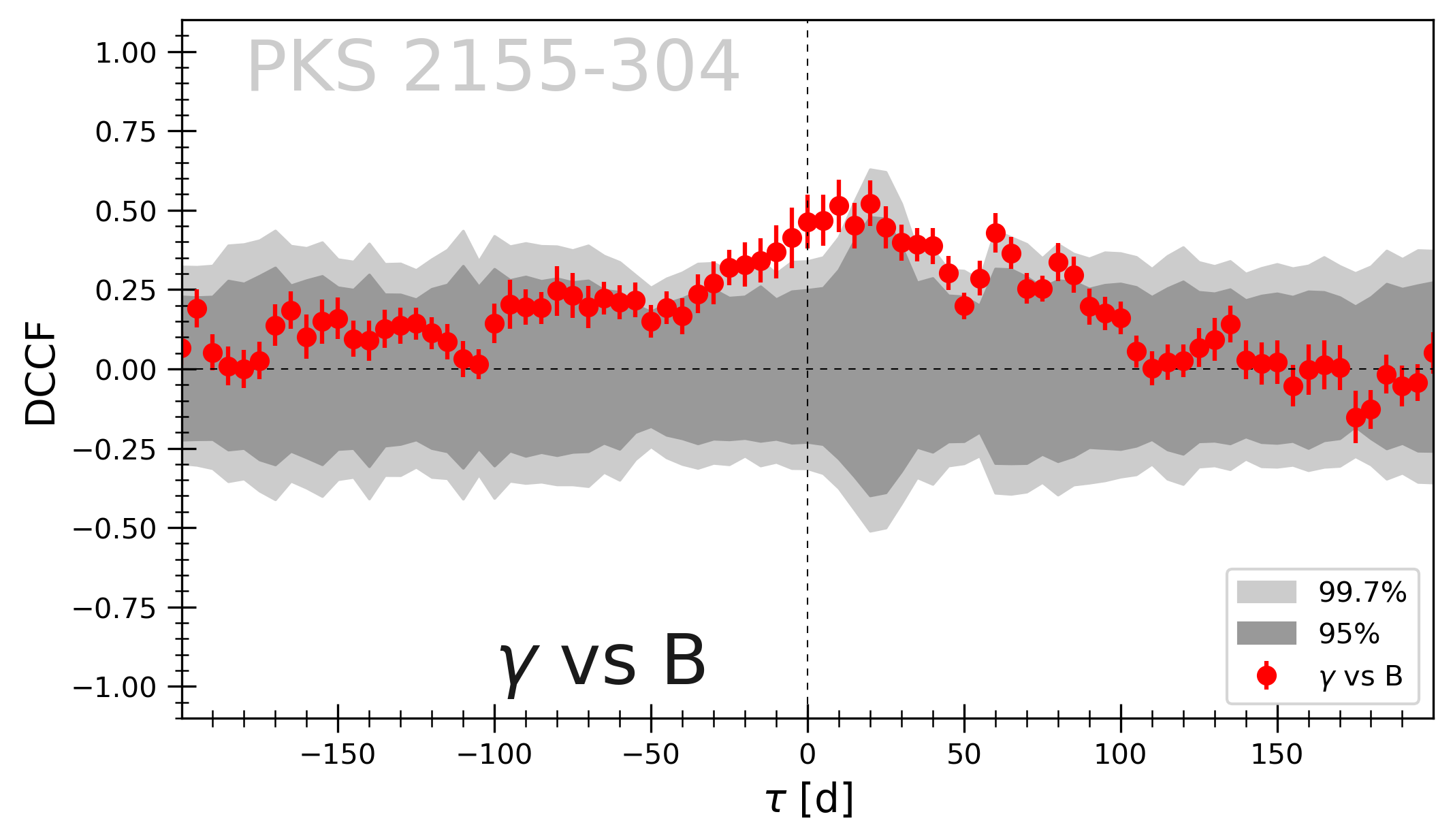}\\
    \includegraphics[width=0.48\textwidth]{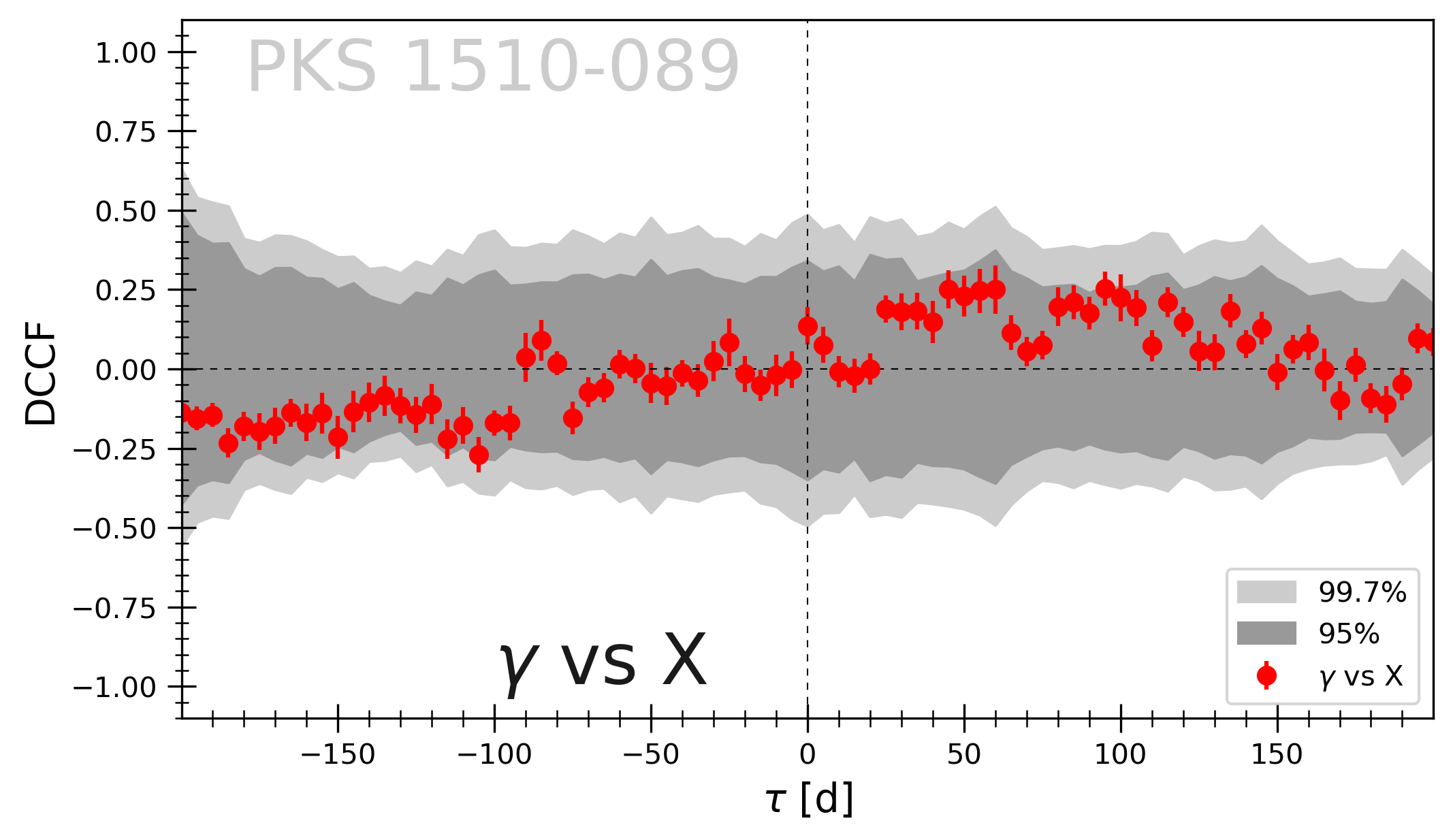}~
    \includegraphics[width=0.48\textwidth]{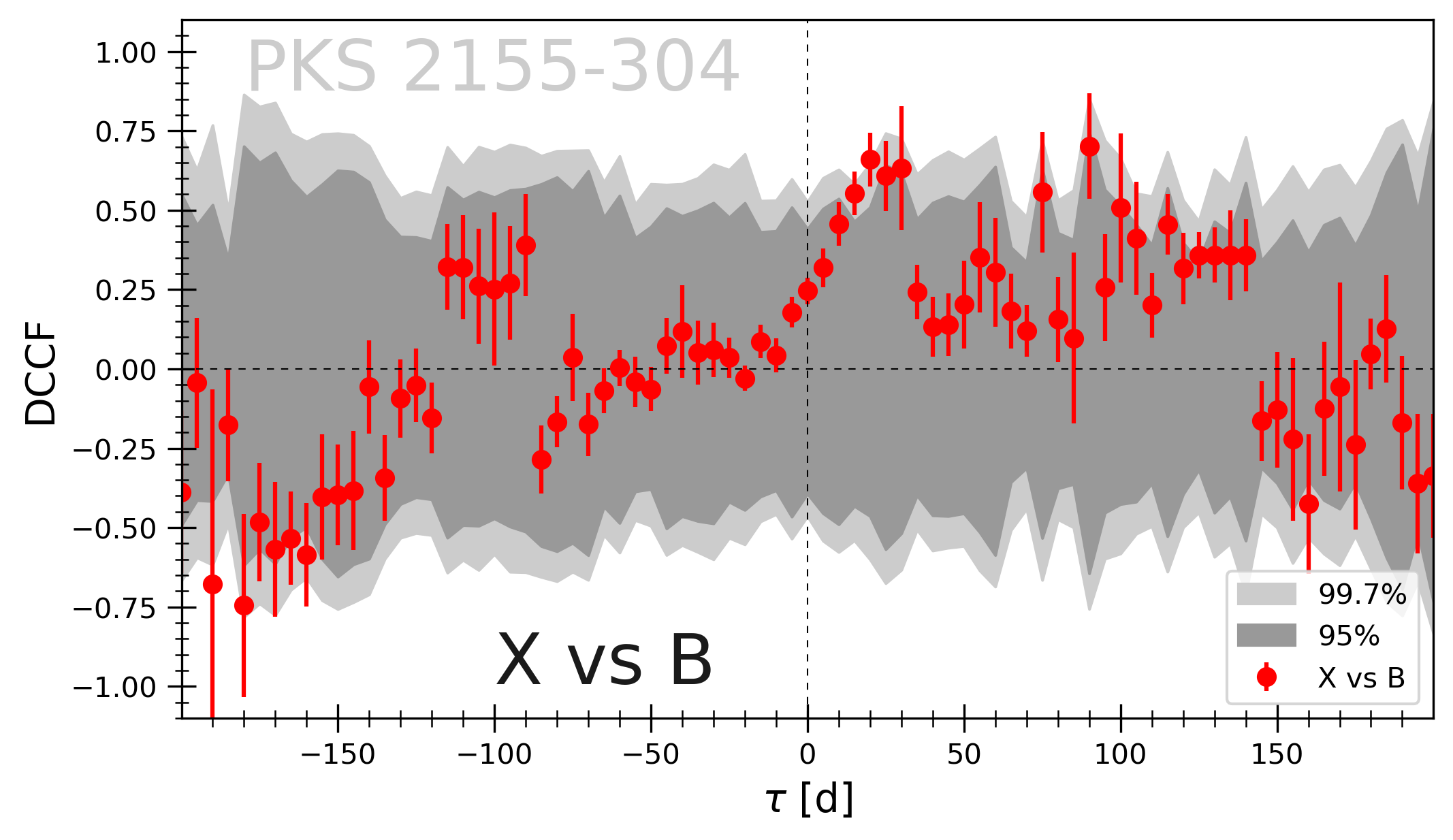}
    \caption{DCCFs of \pksf\ (left) and \pkst (right) showing two sets of correlations each as indicated. The gray bands mark the 2 and 3$\sigma$ contours, respectively.}
    \label{fig:dccf}
\end{figure*}

The discrete cross-correlation function (DCCF) allows one to study the relation between two unevenly-sampled light curves. A Monte Carlo simulation taking into account the power spectral densities and other light curve properties (see \cite{Taylor+26} for details), provides us with significance levels regarding spurious signals. Results are shown in Fig.~\ref{fig:dccf}.

The panels in the top row show the \g-vs-B DCCFs for both sources. As stated above, these bands should trace the same electrons, and thus a high degree of correlation is expected. However, none is visible above the $3\sigma$ level. While the DCCF values for \pkst\ are above this significance level for $\tau\sim0\,$d, they are not at the peak (which is not significant) and they are also at very low DCCF values ($<0.5$). The (non-significant) peaks around $\pm30\,$d and $\pm60\,$d are too close to 1 and 2 months time scales to be considered real.

The bottom row in Fig.~\ref{fig:dccf} shows correlations between bands within the same spectral component. For \pksf, this is the Compton component, while it is the synchrotron component for \pkst. In both cases, one would expect, again, correlated activity. However, no significant correlation is visible.

These results may appear odd, as we know of individual flares or outbursts, where clear correlated activity is visible between these bands. However, these strong correlations may differ from flare to flare including orphan events. Hence, over the course of 20 years the strong correlation of any individual event is washed out by the differing correlation patterns of other events.

\section{Flux distributions}

\begin{figure*}
    \centering
    \includegraphics[width=0.48\textwidth]{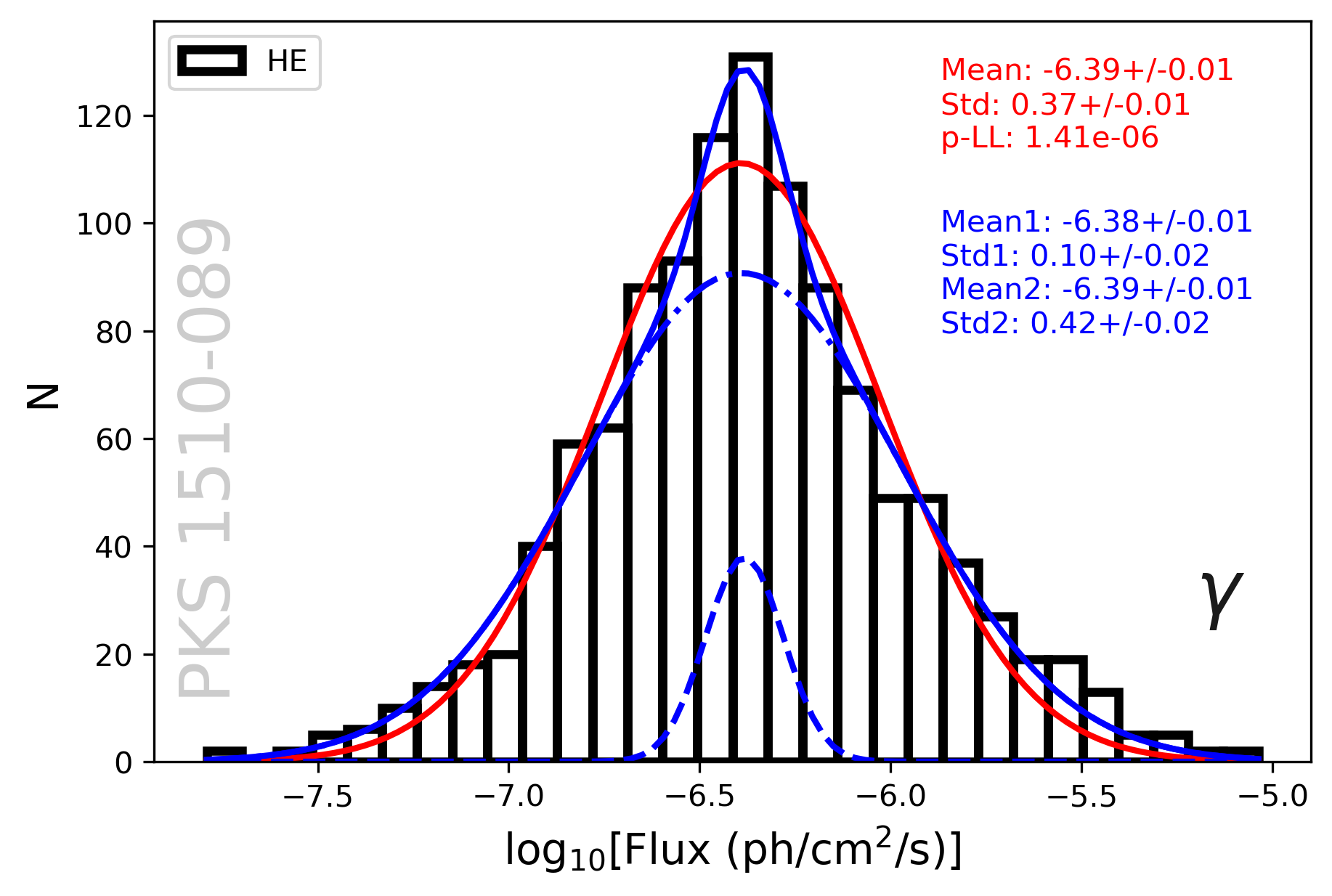}~
    \includegraphics[width=0.48\textwidth]{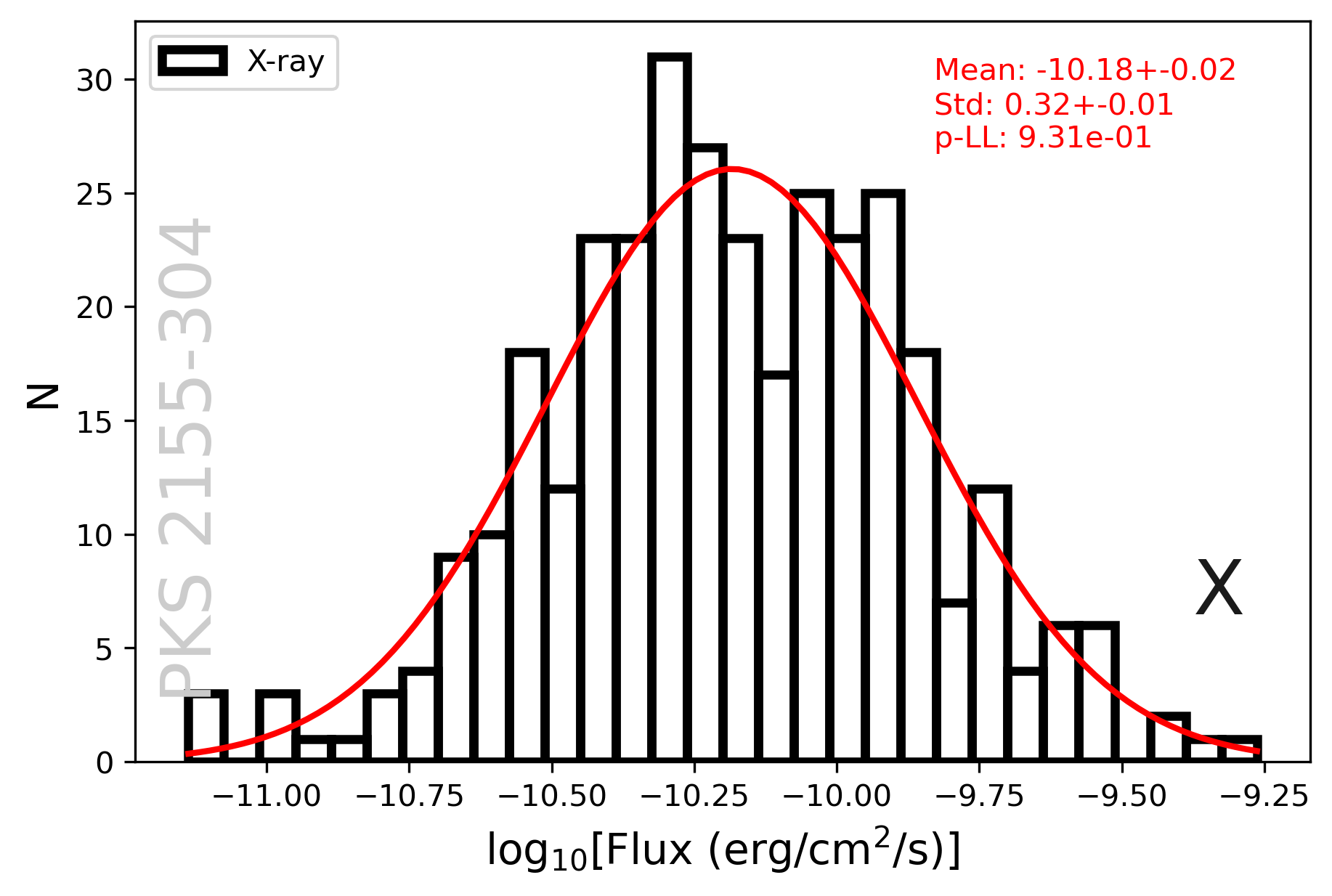}\\
    \includegraphics[width=0.48\textwidth]{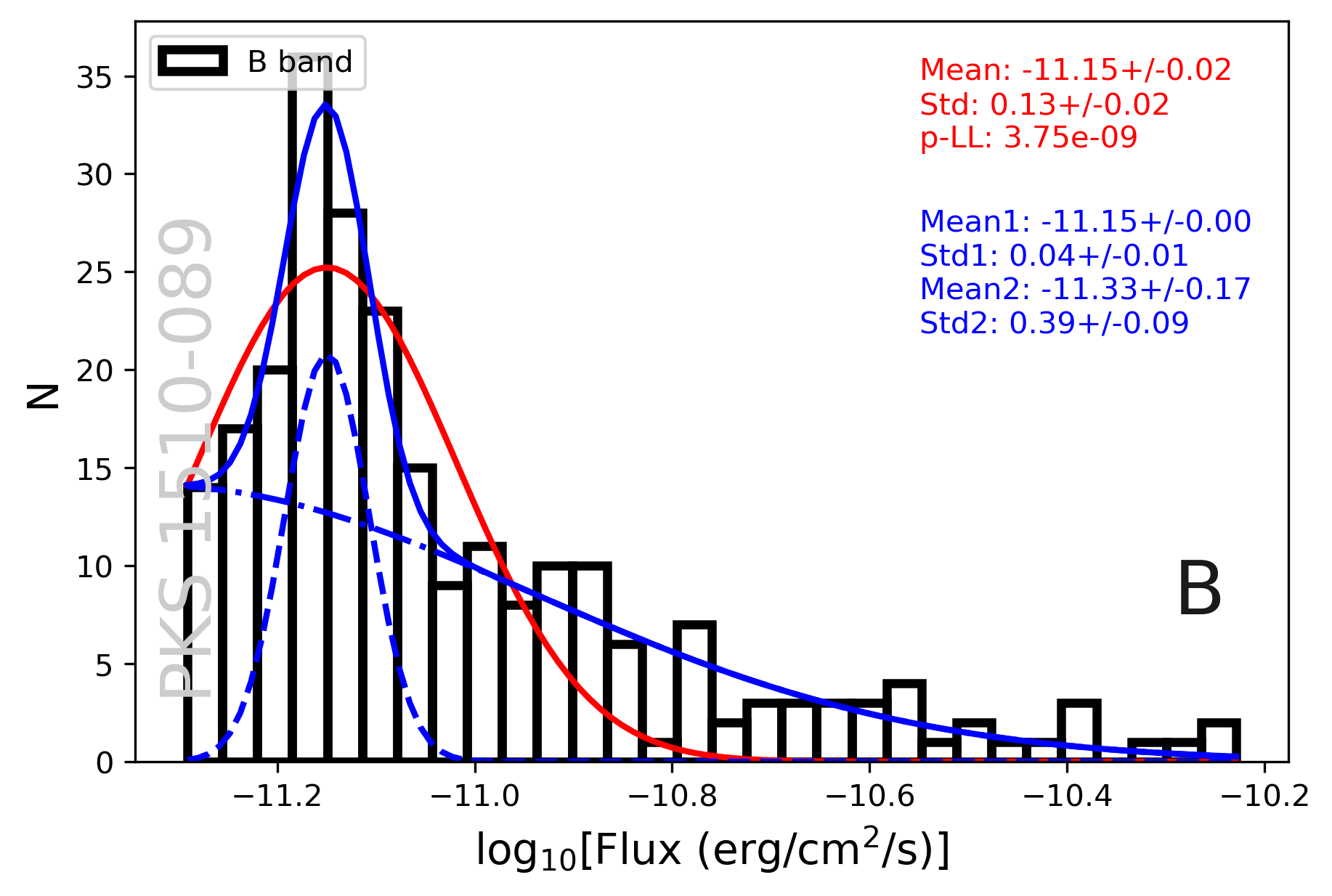}~
    \includegraphics[width=0.48\textwidth]{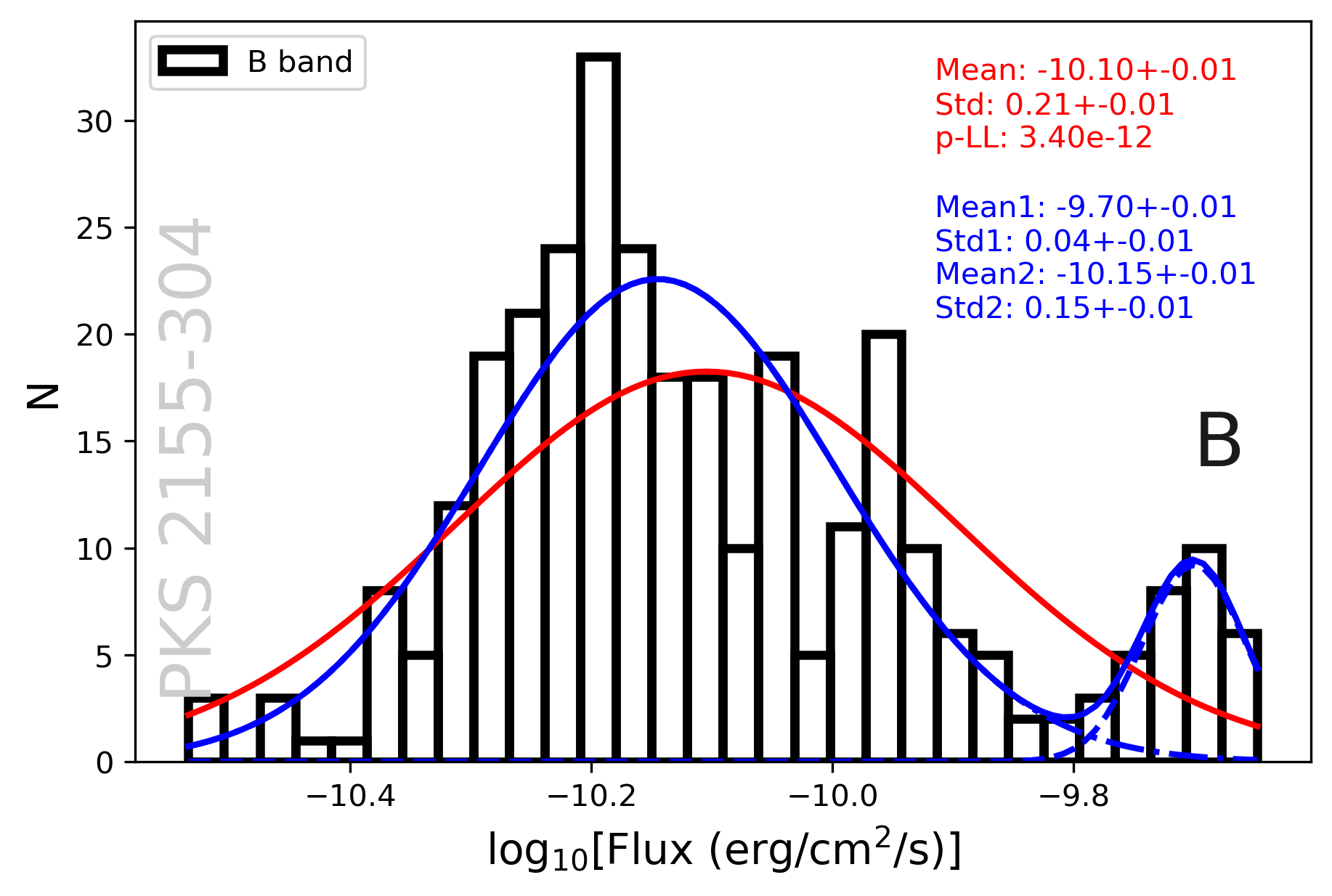}
    \caption{Flux distributions of \pksf\ (left) and \pkst (right) showing two energy bands each as indicated. The red line is a log-normal fit, while the blue lines are a double-log-normal fit (thick: total; dashed and dash-dotted: subcomponents).}
    \label{fig:fluxdist}
\end{figure*}

Flux distributions are an easy measure to display both the range of fluxes attained in a given energy band, as well as the frequency of their occurrence. This is helpful to see how active a source in this band and how often certain activity levels are reached. The shape of flux distributions is often well fitted by a log-normal distribution; sometimes also by a double-log-normal function with well-separated peaks. The latter is interpreted as two separate states for the source.

Flux distributions for \pksf\ and \pkst\ are shown in Fig.~\ref{fig:fluxdist} for two bands, respectively. For \pksf, we show the distributions for the HE and B band, which are both at the high-energy end of the respective SED component. Both distributions are compatible with a double-log-normal fit, even though now clean separation is possible for the HE band. In the B band, we have shown \cite{ZW26} that the low-flux peak is from an accretion-disk-dominated state, while the broad tail describes jet-dominated states. For \pkst, we show the X-ray and B-band distributions, which are from the high-energy and low-energy domain of the synchrotron component. While the B band shows two separate peaks, we could show \cite{WZ26} that the high-flux peak is exclusively filled with data taken before 2010. At this time, a significant flux drop took place in the optical bands, which is then described by the low-flux peak. Hence, both states are compatible with single log-normal distributions -- much like the X-ray band. In turn, it seems that log-normality is a feature in \pkst, but not in \pksf.

We have found that the attained flux range depends on the energy band. Namely, the higher the energy in a given SED component, the broader the distribution. This means that higher energies of a given SED component exhibit stronger variability than lower energies. This is obvious in \pkst, as the X-ray band covers nearly two orders of magnitude, while the B band does not even cover one order of magnitude despite the aforementioned flux drop.

\section{Summary \& Conclusion}

We have studied the long-term variability of \pksf, an FSRQ, and \pkst, an HBL, using 20 years of data from the \textit{Swift} and \textit{Fermi} instruments.
The variability of both sources shows a couple of similarities. Foremost, we have found strong variability in all considered energy bands, while the range of attained fluxes depends on the position of the observed band within the blazar's SED. At higher energies within an SED component, the flux range is broader than at lower energies of the same SED component. Strikingly, we have found no global correlations between the observed bands implying changing correlation patterns from flare to flare.

We have also found notable differences between the sources. For instance, \pkst\ flux is apparently log-normally distributed, while this is not the case in \pksf. Additionally, as described in detail in \cite{WZ26,ZW26}, \pkst\ shows harder-when-brighter behaviour in its X-ray domain (high-energy part of the synchrotron component), even though the actual track differs from year to year. No such correlation is found in \pksf\ irrespective of the energy band. On the other hand, the year-wise fractional variability correlates with the year-wise average flux in the HE and B band in \pksf, but there is no correlation in any band in \pkst.

None of these findings appear consistent with the expectations of the standard one-zone model. This is to say, it appears unlikely that it is always just the same zone that is contributing to the observed flux. Multi-zone or stratified-jet models \cite{zacharias+22} are more likely to fully capture the long-term evolution in blazars. We already know that \pksf\ must have had at least two active zones in the past \cite{hess23}. It seems logical that many if not most other blazars should also be governed by more than one emission region \cite{zachariaswagner16,hess20}.

The similarities found between these two sources nevertheless suggest that the underlying physical processes in their jets are comparable irrespective of the source type. And it has also become obvious that the observed variability is, in fact, variable.

\printbibliography[title={References}]

%
%
%

\end{document}